\documentclass[showpacs,aps,twocolumn]{revtex4}
\usepackage{epsf}
\newcommand{\mb}[1]{ { \mbox{\boldmath{$#1$}}}  } 
 
\begin{document}
    
\title{In-gap states of the quantum dot coupled between a
       normal and superconducting lead}

\author{J.\ Bara\'nski  and  T.\ Doma\'nski}
\affiliation{
       Institute of Physics, M.\ Curie-Sk\l odowska University, 
       20-031 Lublin, Poland}

\date{\today}

\begin{abstract}
We study the in-gap states of the quantum dot hybridized with one 
conducting and another superconducting electrode. Proximity effect 
suppresses the electronic states in the entire subgap regime $|\omega| 
< \Delta$, where $\Delta$ denotes the energy gap of a superconductor.  
The Andreev scattering mechanism can induce, however, some in-gap 
states of a line-broadening (inverse life-time) controlled by a 
hybridization of the quantum dot with a normal electrode. We show that 
the number of such Andreev bound states is substantially dependent
on a competition between the Coulomb repulsion and the induced 
on-dot pairing.  We discuss signatures of these in-gap states 
in the tunneling conductance, especially in a low-bias regime.  
\end{abstract}

\maketitle

\section{Motivation}

The quantum impurities embedded in superconducting host materials have been the
topic of intensive studies for about 50 years (see the review paper \cite{Balatsky-06}). 
Early interests predominantly explored in what way the impurities affect 
the superconducting state of the bulk materials. It has been established (by 
the Anderson theorem \cite{Anderson-59}) that the paramagnetic impurities 
are rather inefficient on the isotropic superconductors or eventually 
weakly suppress the anisotropic superconducting phases \cite{Abrikosov-61}. 
The magnetic (spinful) impurities, on contrary,  proved to have much stronger 
influence on superconductivity. They induce the in-gap states \cite{Shiba-Rusinov} 
and with increasing concentration of the magnetic impurities the energy gap of 
superconducting material is gradually filled-in simultaneously suppressing  
its critical temperature. On a microscopic level this detrimental effect  
comes from a pair-breaking character of the spin scattering. 

Mutual relationship between the quantum impurities and super\-conducting materials 
attracts presently again substantial interests due to intensive studies of the nanoscopic 
devices, where various artificial quantum impurities (dots) are connected to 
the external superconducting electrodes. In this context, the main problem
refers to the question {\em how do the superconducting reservoirs affect 
the quantum dots} rather than the other way around. Due to the proximity 
effect the Cooper pairs can penetrate the quantum dot, converting it into 
a sort of the 'superconducting grain'. On the other hand, the strong Coulomb repulsion 
between opposite spin electrons disfavors any double (or even) occupancy of 
the quantum dot.  At low temperatures also the Kondo physics additionally 
comes into the play. Both these phenomena, i.e.\ the Coulomb blockade and 
appearance of the Kondo singlet state, strongly compete with the induced 
on-dot pairing. In nanoscopic tunneling junctions this competition can 
be explored  in a controllable manner, by: a) varying the QD hybridization 
with the superconducting lead, b) affecting the energy gap $\Delta
=\Delta(B)$  by applying the magnetic field $B$ \cite{DeFranceschi-12}, 
c) lifting the discrete QD energy levels via the gate voltage, and 
d) lowering temperatures to activate the Kondo physics. 
Numerous theoretical and experimental studies of the quantum dots 
connected to the superconducting leads have been summarized 
e.g.\ in Refs \cite{Rodero-11,DeFranceschi-10}.  

Interplay between the on-dot pairing and the correlation effects can be conveniently
investigated in the setup, where the quantum dots are placed between one 
superconducting (S) and another normal (N) electrode. In the subgap regime ($eV < \Delta$) 
the tunneling conductance practically entirely originates from the anomalous 
Andreev channel,  such spectroscopy can thus directly probe any in-gap 
states. In experimental realizations of the N-QD-S junctions the role of quantum dots
has been played by the self-assembled InAs nanoscopic islands \cite{Deacon-10}, 
carbon nanotubes \cite{Pillet-10}, quantum wires \cite{Kouwenhoven-12} etc. 
For instance, using InAs quantum dots coupled between the metallic (golden) 
and superconducting (aluminum) electrodes provided a clear evidence for the Kondo 
effect coexisting with the induced on-dot pairing manifested by the zero-bias 
enhancement of the zero-bias Andreev conductance \cite{Deacon-10}. Tunneling 
conductance has been recently measured also in the system comprising the indium 
antimonide nanowires connected to a normal (gold) and superconducting (niobium 
titanium nitride) electrode, indicating the Majorana type in-gap states 
\cite{Kouwenhoven-12}. 

Other measurements have been done using the three-terminal configurations
with the metallic and superconducting electrodes interconnected via the double 
quantum dots to achieve a controllable Cooper pair splitting. These dots served 
as 'quantum forks', where the Coulomb repulsion enforced electrons (released from 
the Cooper pairs) to move into different normal leads, yet preserving their 
entanglement. Such transport channel contributed about 10 percent (for the case
of InAs quantum dots \cite{Hofstetter-09}) and nearly 50 percent (using the 
carbon nanotubes \cite{Hermann-10}) to the total differential conductance. 
For the latter case the efficiency has been next considerably improved \cite{Schindele-12}. 
In experimental measurements there have also probed the spin-polarized Andreev current
using the ferromagnetic electrode coupled via the quantum dots to the superconducting 
lead \cite{Hofstetter-10}. In all these and many other related experiments 
\cite{S-QD-S,S-QD_and_N-S} the subgap electron transport  is solely provided 
by the in-gaps states. Their detailed knowledge seems thus to be a timely 
and important issue.
 
It is our intention here to gather a systematic information on the in-gap Andreev 
states originating from the scattering either on the magnetic or non-magnetic quantum 
impurities. Physical aspects of such study have been so far addressed by a number of 
groups using various techniques \cite{Rodero-11}. Since this problem is presently 
important \cite{Aguado-13,Viewpoint} we would like to collect the essential results 
into this single report. Subgap states of the magnetic (Kondo-type) impurities 
have been extensively investigated, both theoretically \cite{Koerting-10} and 
experimentally using the two-terminal \cite{Lohneysen-12} as well as three-terminal 
configurations \cite{S-QD_and_N-S}. We would like to emphasize, however, that
in-gap states are present also in the case of uncorrelated (spinless) quantum 
dots \cite{Bauer-07,Hecht-08,Beenakker-92}. To illustrate such possibility in 
section III we briefly analyze  the noninteracting quantum dot, considering 
evolution of the Andreev bound states with respect to  
$\Delta/\Gamma_{S}$ [where $\Gamma_{\beta}$ denotes the coupling to $\beta=N,S$ 
lead] for several asymmetric coupling ratios $\Gamma_{S}/\Gamma_{N}$. Next, 
in sections IV and V, we address the correlation effects responsible for 
the Coulomb blockade and the Kondo effect. 

Our study can be regarded as complementary to the previous pedagogical 
analysis by J.\ Bauer {\em et al} \cite{Bauer-07} who focused on the 
in-gap states of the quantum impurity immersed in a superconducting 
medium for the limit $\Gamma_{S}\gg \Delta$. We hope that this 
analysis would be useful for the tunneling spectroscopy using  
the quantum dots asymmetrically coupled between the superconducting 
and normal leads in the two- and multi-terminal configurations.

\section{Anderson impurity model}

For description of the quantum dot coupled between the normal (N) and 
superconducting (S) electrodes we use the Anderson impurity model 
\begin{eqnarray} 
\hat{H} &=& \hat{H}_{N} + \hat{H}_{S} + \sum_{\sigma}  
\epsilon_{d} \hat{d}^{\dagger}_{\sigma} \hat{d}_{\sigma}  
+  U_{d} \; \hat{n}_{d \uparrow} \hat{n}_{d \downarrow}  
\nonumber \\
&+& \sum_{{\bf k},\sigma } \sum_{{\beta}=N,S}  
\left( V_{{\bf k} \beta} \; \hat{d}_{\sigma}^{\dagger}  
\hat{c}_{{\bf k} \sigma \beta } + V_{{\bf k} \beta}^{*}  
\; \hat{c}_{{\bf k} \sigma, \beta }^{\dagger} \hat{d}_{\sigma} 
\right) . 
\label{model} 
\end{eqnarray} 
Operators $d_{\sigma}$ ($d_{\sigma}^{\dagger}$) denote annihilation  
(creation) of QD electron with spin $\sigma$ and energy level 
$\varepsilon_{d}$ and $U_{d}$ is the on-dot repulsion (or charging) energy. 
The last term in (\ref{model}) represents a hybridization of the QD with 
the external leads, where the normal electrode is described by the Fermi 
gas $\hat{H}_{N} \!=\! \sum_{{\bf k},\sigma} \xi_{{\bf k}N}  
\hat{c}_{{\bf k} \sigma N}^{\dagger} \hat{c}_{{\bf k} \sigma N}$  
and the superconducting one is takes the conventional BCS form
$\hat{H}_{S} \!=\!\sum_{{\bf k},\sigma}  \xi_{{\bf k}S}\hat{c}_{{\bf k} 
\sigma S }^{\dagger}  \hat{c}_{{\bf k} \sigma S} \!-\! \sum_{\bf k} 
\Delta  \left( \hat{c}_{{\bf k} \uparrow S }^{\dagger} \hat{c}_{-{\bf k} 
\downarrow S }^{\dagger} + \hat{c}_{-{\bf k} \downarrow S} \hat{c}_{{\bf k} 
\uparrow S }\right)$. The energies $\xi_{{\bf k}\beta}\!=\!\varepsilon_{
{\bf k}\beta} \!-\!\mu_{\beta}$ are measured with respect to the chemical 
potentials $\mu_{\beta}$, which can be detuned by the external voltage 
$\mu_{N}\!=\!\mu_{S}\!+\!eV$. We shall focus on the low energy features, 
assuming the wide band limit approximation $|V_{{\bf k}\beta}| \!  \ll \! D$ 
(where $-D\!\leq\!\varepsilon_{{\bf k}\beta} \!  \leq \! D$) and
use the coupling constants $\Gamma_{\beta}=2\pi\sum_{{\bf k},\beta}
|V_{{\bf k}\beta}|^{2}\delta(\omega-\xi_{{\bf k}\beta})$ as useful
energy units.

To consider the proximity effect we introduce the matrix Green's function 
${\mb G}_{d}(\tau,\tau')\!=\!\langle\langle \hat{\Psi}_{d}(\tau); \hat{\Psi}_{d}^{\dagger}
(\tau')\rangle\rangle$ in the Nambu representation $\hat{\Psi}_{d}^{\dagger}=
(\hat{d}_{\uparrow}^{\dagger},\hat{d}_{\downarrow})$, $\hat{\Psi}_{d}=
(\hat{\Psi}_{d}^{\dagger})^{\dagger}$. Under equilibrium conditions the Green's 
function ${\mb G}_{d}(\tau,\tau')$ depends only on a time difference $\tau\!
-\!\tau'$. Its Fourier transform obeys the following Dyson equation
\begin{eqnarray} 
{\mb G}_{d}(\omega)^{-1} = 
\left( \begin{array}{cc}  
\omega\!-\!\varepsilon_{d} &  0 \\ 0 &  
\omega\!+\!\varepsilon_{d}\end{array}\right)
- {\mb \Sigma}_{d}^{0}(\omega)  
- {\mb \Sigma}_{d}^{U}(\omega) ,  
\label{GF}\end{eqnarray} 
where the selfenergy ${\mb \Sigma}_{d}^{0}$ corresponds to the noninteracting 
case ($U\!=\!0$) and the second contribution ${\mb  \Sigma}_{d}^{U}$ refers 
to the correlation effects induced by local Coulomb repulsion $U_{d}\hat{n}
_{d,\uparrow}\hat{n}_{d,\downarrow}$. The uncorrelated quantum dot is characterized
by 
\begin{eqnarray}
\mb{\Sigma}_{d}^{0}(\omega)=\sum_{{\bf k}, \beta} \left|V_{{\bf k} \beta} \right|^{2}
\; \mb{g}_{\beta}({\bf k}, \omega) ,
\label{Sigma_0_QD}
\end{eqnarray}
where ${\mb g}_{N}({\bf k}, \omega)$ is the Green's function of the normal lead 
\begin{eqnarray} 
{\mb g}_{N}({\bf k}, \omega) = 
\left( \begin{array}{cc}  
\frac{1}{\omega-\xi_{{\bf k}N}} & 0 \\ 
0 &  
\frac{1}{\omega+\xi_{{\bf k}N}}
\end{array}\right)
\label{gN}
\end{eqnarray} 
and ${\mb g}_{S}({\bf k}, \omega)$ denotes the Green's function of superconducting electrode
\begin{eqnarray} 
{\mb g}_{S}({\bf k}, \omega) = 
\left( \begin{array}{cc}  
\frac{u^{2}_{\bf k}}{\omega-E_{\bf k}}+\frac{v^{2}_{\bf k}}
{\omega+E_{\bf k}} \hspace{0.2cm} & \frac{-u_{\bf k}v_{\bf k}}
{\omega-E_{\bf k}}+\frac{u_{\bf k}v_{\bf k}}{\omega+E_{\bf k}}
\\ 
\frac{-u_{\bf k}v_{\bf k}}{\omega-E_{\bf k}}+
\frac{u_{\bf k}v_{\bf k}}{\omega+E_{\bf k}}
& \frac{u^{2}_{\bf k}}{\omega+E_{\bf k}}+
\frac{v^{2}_{\bf k}}{\omega-E_{\bf k}}
\end{array}\right) .
\label{gS}\end{eqnarray} 
The quasiparticle energies are given by
$E_{\bf k}\!=\!\sqrt{\xi_{{\bf k}S}^{2}+\Delta^{2}}$ and 
the usual BCS coefficients take a form 
$u^{2}_{\bf k},v^{2}_{\bf k} = \frac{1}{2} \left[ 1 \pm 
\frac{\xi_{{\bf k}S}}{E_{\bf k}} \right]$,
$u_{\bf k}v_{\bf k} = \frac{\Delta}{2E_{\bf k}}$.
In the wide-band limit the selfenergy (\ref{Sigma_0_QD}) simplifies to 
\begin{eqnarray}
\mb{\Sigma}_{d}^{0}(\omega) =  -i \frac{\Gamma_{N}}{2} \; 
\left( \begin{array}{cc}  
1 & 0 \\ 0 & 1 \end{array} \right)
- \frac{\Gamma_{S}}{2} \gamma(\omega)
\left( \begin{array}{cc}  
1 & \frac{\Delta}{\omega} \\ 
 \frac{\Delta}{\omega}  & 1 
\end{array} \right)
\label{selfenergy_0}
\end{eqnarray} 
with $\omega$-dependent function
\begin{eqnarray}
\gamma(\omega) = \left\{
\begin{array}{ll} 
\frac{\omega}{\sqrt{\Delta^{2}-\omega^{2}}}
& \mbox{\rm for }  |\omega| < \Delta , \\
\frac{i\;|\omega|}{\sqrt{\omega^{2}-\Delta^{2}}}
& \mbox{\rm for }  |\omega| > \Delta .
\end{array} \right.
\label{gamma}
\end{eqnarray} 
For considering the correlation effects $\Sigma_{d}^{U}(\omega)$ 
one has to introduce some approximations. We shall come back to 
this non-trivial problem in sections IV and V.

\begin{figure}
\epsfxsize=12cm\centerline{\epsffile{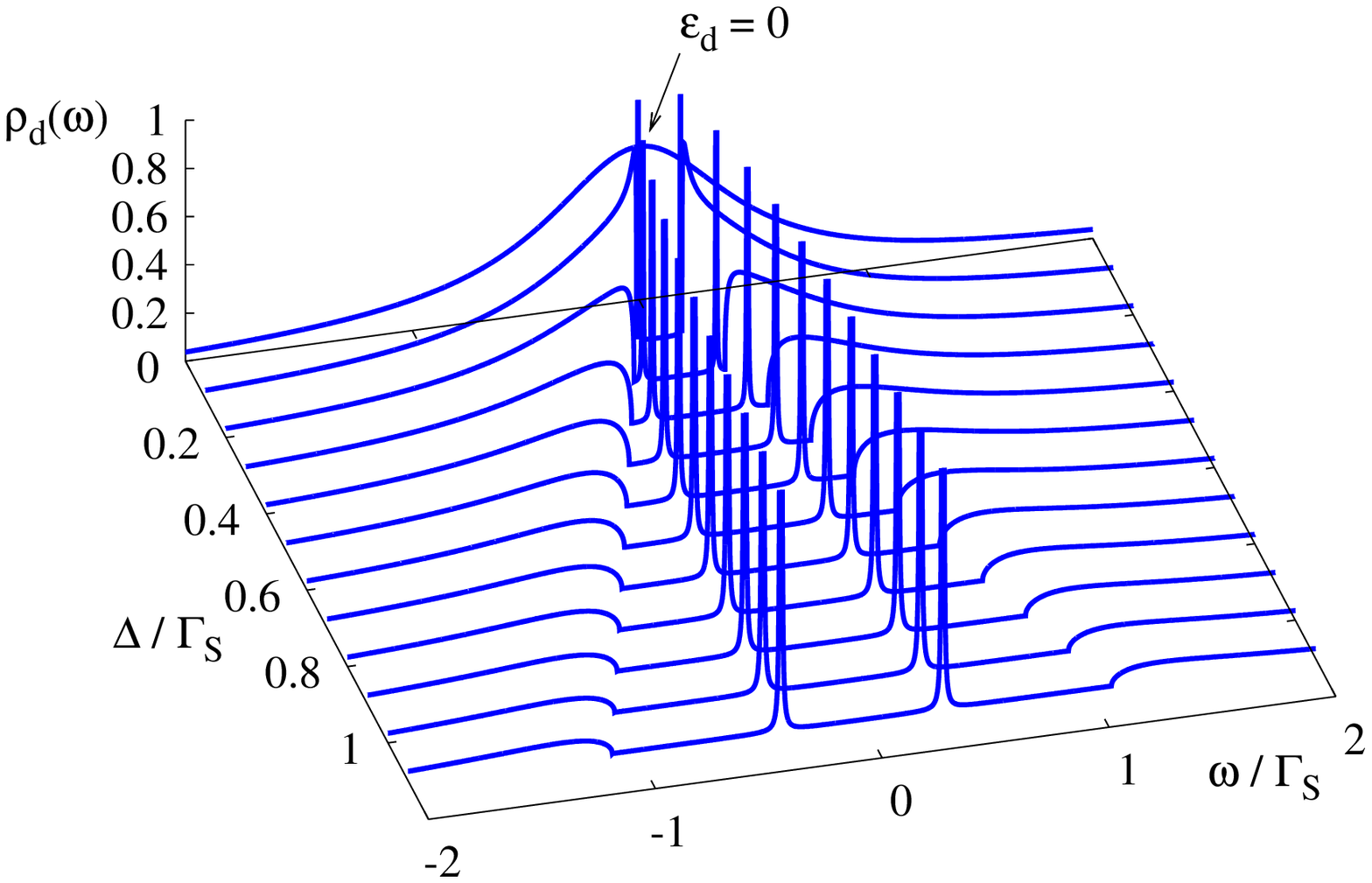}}
\vspace{-1.0cm}
\epsfxsize=12cm\centerline{\epsffile{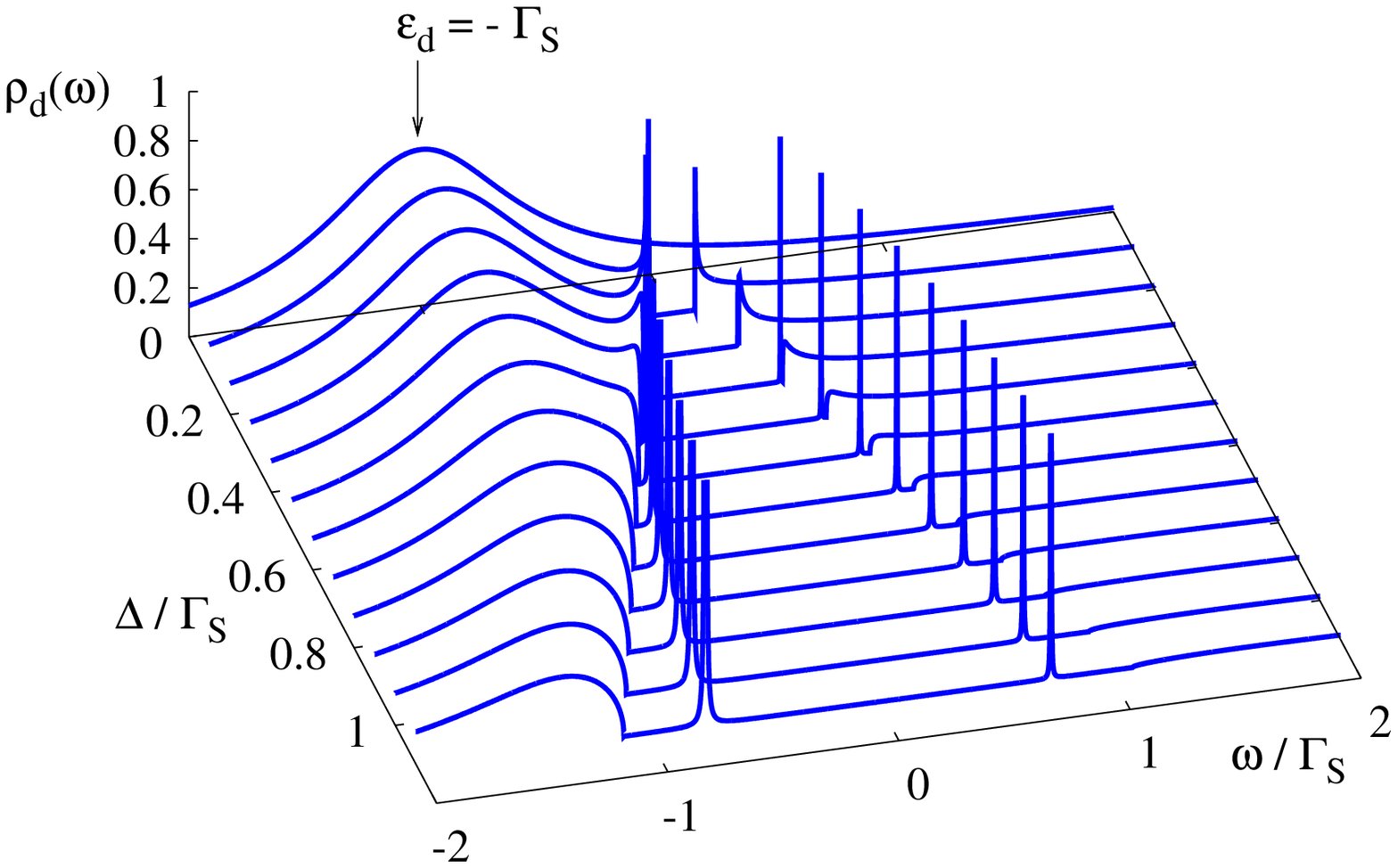}}
\vspace{0.0cm}
\caption{(color online) Spectral function $\rho_{d}(\omega)$ of 
the uncorrelated quantum dot obtained for $\varepsilon_{d}=0$ (upper panel) 
and $\varepsilon_{d}=-\Gamma_{S}$ (bottom panel) assuming weak coupling
to the metallic lead $\Gamma_{N}=0.001\Gamma_{S}$. In both cases 
the in-gap states gradually emerge from the gap edge singularities $\pm \Delta$ 
(when $\Delta \ll \Gamma_{S}$) and they evolve to the well-defined subgap 
quasiparticle peaks (when $\Delta \gg \Gamma_{S}$).}
\label{proximity_effect}
\end{figure}

\section{In-gap states of the uncorrelated quantum dot}

Let us start by considering the uncorrelated QD, which is equivalent to the 
spinless impurity. We discuss here the spectroscopic properties of such QD 
for an arbitrary ratio $\Delta/\Gamma_{S}$ and asymmetric couplings $\Gamma_{N} 
\neq \Gamma_{S}$. In-gap states formally represent the poles of the matrix 
Green's function ${\mb G}_{d}(\omega)$ existing in the subgap regime $|\omega| 
< \Delta$. For the uncorrelated quantum dot ($U_{d}\rightarrow 0$) one has
\begin{eqnarray}
{\mb G}_{d}(\omega) &=& \frac{1}{\left( \tilde{\omega} + i 
\frac{\Gamma_{N}}{2}\right)^{2} - \varepsilon_{d}^{2} - 
\left( \frac{\tilde{\Gamma}_{S}}{2} \right)^{2}} \; 
\nonumber \\ & \times &
\left( \begin{array}{cc}  
\tilde{\omega} + i \frac{\Gamma_{N}}{2} + \varepsilon_{d} 
\hspace{0.3cm}& -\; \frac{\tilde{\Gamma}_{s}}{2} \\ 
 -\; \frac{\tilde{\Gamma}_{s}}{2}  & 
\tilde{\omega} + i \frac{\Gamma_{N}}{2} - \varepsilon_{d} 
\end{array} \right)
\label{G_0}
\end{eqnarray}
with the following meaning of the symbols $\tilde{\omega}$ 
and $\tilde{\Gamma}_{S}$ 
\begin{eqnarray}
\tilde{\omega} & = & \omega + \frac{\Gamma_{S}}{2} \frac{\omega}
{\sqrt{\Delta^{2}-\omega^{2}}} , \label{tilde_omega} \\
\tilde{\Gamma}_{s} & = & \Gamma_{S} \frac{\Delta}
{\sqrt{\Delta^{2}-\omega^{2}}} .
\label{tilde_GammaS}
\end{eqnarray}
In this case the single-particle spectral function  $\rho_{d}(\omega) 
\equiv -\;\frac{1}{\pi} \mbox{\rm Im}\left\{ {\mb G}_{d,11}(\omega) \right\}$  
is expressed  by the standard BCS form
\begin{eqnarray}
{\mb G}_{d,11}(\omega) &=& \frac{1}{2} \left[ 1 + \frac{\varepsilon_{d}}
{\tilde{E}_{d}} \right] \frac{1}{ \tilde{\omega} + i 
\frac{\Gamma_{N}}{2} - \tilde{E}_{d}} 
\\ \nonumber
&+& \frac{1}{2} \left[ 1 - \frac{\varepsilon_{d}}
{\tilde{E}_{d}} \right] \frac{1}{ \tilde{\omega} + i 
\frac{\Gamma_{N}}{2} + \tilde{E}_{d}}
\label{G11_0}
\end{eqnarray}
with $\omega$-dependent parameter
\begin{eqnarray}
\tilde{E}_{d} = \sqrt{ \varepsilon_{d}^{2} + \left(
\tilde{\Gamma}_{s}/2 \right)^{2}} .
\label{tilde_Ed} 
\end{eqnarray}
In figure \ref{proximity_effect} we show the spectrum $\rho_{d}(\omega)$ 
as a function of the energy gap $\Delta$ obtained for the uncorrelated 
(spinless) quantum dot  with $\varepsilon_{d}=0$ and $\varepsilon_{d}
=-\Gamma_{S}$. We have assumed a weak coupling to the normal lead 
$\Gamma_{N} \ll \Gamma_{S}$ what yields a nearly resonant character 
of the in-gap states. For larger $\Gamma_{N}$ the in-gap states 
broadening increases (life-time decreases). We furthermore notice 
(see Fig.\ \ref{in-gap-energies}),
that the Andreev states appear near the gap edge singularities 
(for $\Delta \ll \Gamma_{S}$) and they evolve into the subgap peaks 
centered at energies $\pm\sqrt{\varepsilon_{d}^{2}+\Gamma_{S}^{2}/4}$ 
(for $\Delta \gg \Gamma_{S}$). 

\subsection{Resonances in the weak coupling limit 
             \boldmath{$\Gamma_{N} \rightarrow 0$}}

\begin{figure}
\epsfxsize=9cm\centerline{\epsffile{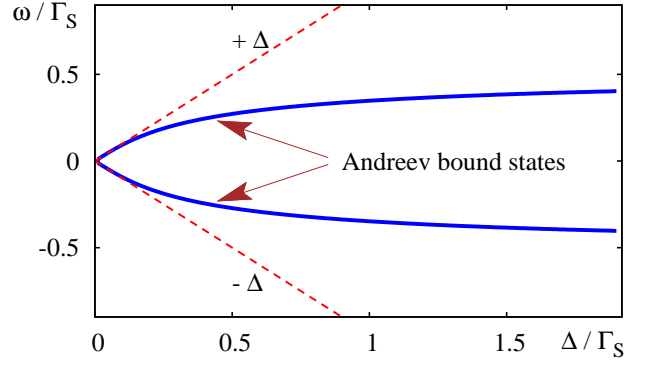}}
\vspace{0.0cm}
\caption{(color online) Energies of the in-gap states versus the ratio 
$\Delta/\Gamma_{S}$ obtained for the uncorrelated quantum dot ($\varepsilon_{d}=0$)
weakly coupled to the metallic lead $\Gamma_{N} = 0.001 \Gamma_{S}$. 
Dashed lines indicate the gap edges $\pm \Delta$.}
\label{in-gap-energies}
\end{figure}

To get some correspondence with the previous study \cite{Bauer-07} we now
consider in more detail the case of infinitesimally weak coupling to the normal electrode
$\Gamma_{N} \rightarrow 0^{+}$. Under such condition the in-gap states become
strictly resonant, i.e.\ they represent the quasiparticles
of an infinite life-time. In the subgap regime the Eqn (\ref{G11_0}) yields 
the following spectral function 
\begin{eqnarray}
\lim_{|\omega|<\Delta} \rho_{d}(\omega) &=& \frac{1}{2} \left( 1 + \frac{\varepsilon_{d}}
{\tilde{E}_{d}} \right) \delta \left[ \tilde{\omega} - \tilde{E}_{d}\right] 
\nonumber \\ & + &
\frac{1}{2} \left( 1 - \frac{\varepsilon_{d}}
{\tilde{E}_{d}} \right) \delta \left[ \tilde{\omega} + \tilde{E}_{d}\right] .
\label{resonant_spectrum}
\end{eqnarray}
This function (\ref{resonant_spectrum}) can be rewritten as
\begin{eqnarray}
\lim_{|\omega|<\Delta} \rho_{d}(\omega) = {\cal{W}}_{+} \; \delta \left[ \omega 
- E_{+} \right] + {\cal{W}}_{-} \; \delta \left[ \omega - E_{-} \right] 
\nonumber \\
\label{resonant_spectrum_bis}
\end{eqnarray}
with the quasiparticle energies $E_{\pm}$ representing the solutions of 
the following equation
\begin{eqnarray}
E_{\pm} + \frac{\Gamma_{S}}{2} \frac{E_{\pm}} {\sqrt{\Delta^{2}
-E_{\pm}^{2}}} = \pm \sqrt{\varepsilon_{d}^{2}+\left( \frac{
\Gamma_{S}}{2}\right)^{2} \frac{\Delta^{2}}{\Delta^{2}-E_{\pm}^{2}}} .
\nonumber \\
\label{energy_eqn}
\end{eqnarray}
and ${\cal{W}}_{\pm}$ being their spectral weights.

We illustrate in figure \ref{in-gap-energies} the energies $E_{\pm}$ of 
the in-gap resonances versus the ratio $\Delta/\Gamma_{S}$ obtained for 
$\varepsilon_{d}=0$. In the case of small energy gap $\Delta \ll 
\Gamma_{S}$ (studied by J.\ Bauer {\em et al} \cite{Bauer-07}) the 
resonant in-gap states are located nearby the gap edge singularities 
$\pm\Delta$. For increasing $\Delta/\Gamma_{S}$ they gradually move 
aside from the gap edge singularities, and in the limit $\Delta \gg 
\Gamma_{S}$ approach the asymptotic values $\pm \sqrt{\varepsilon_{d}
+\left(\Gamma_{S}/2\right)^{2}}$. In the next section we discuss in
some more detail this 'superconducting atomic' limit $\Delta \gg \Gamma_{S}$.

\subsection{Superconducting atomic limit 
             \boldmath{$\Delta \gg \Gamma_{S}$}}

Deep inside the energy gap (i.e.\ for $| \omega| \ll \Delta$) all electronic 
states of the uncorrelated quantum dot can be determined analytically (for 
arbitrary  $\Gamma_{\beta}$) due to the fact, that the selfenergy 
(\ref{selfenergy_0}) simplifies then to a static value
\begin{eqnarray} 
{\mb \Sigma}_{d}^{0}(\omega) =  - \; \frac{1}{2}
\left( \begin{array}{cc}  i \Gamma_{N} & \Gamma_{S} \\  
\Gamma_{S} & i \Gamma_{N}   
\end{array} \right)  .
\label{Sigma0}
\end{eqnarray} 
Under such conditions the quantum dot can be regarded as the 'superconducting 
island' with the induced pairing gap $\Delta_{d}=|\Gamma_{S}/2|$. This 
problem has been widely discussed in the literature adopting various methods 
to describe the correlation effects $U_{d}$  (see sections IV $\&$ V). 

The spectral function $\rho_{d}(\omega)$ of the uncorrelated QD can be 
expressed explicitly by
\begin{eqnarray} 
\rho_{d}(\omega) &=& \frac{1}{2} \left[ 1 + 
\frac{\varepsilon_{d}}{E_{d}} \right]  \frac{\frac{1}{\pi} \;  
\Gamma_{N}/2}{(\omega\!-\!E_{d})^{2}+ 
(\Gamma_{N}/2)^{2}} \nonumber \\ &+& \frac{1}{2}  
\left[ 1 - \frac{\varepsilon_{d}}{E_{d}} \right]  
\frac{\frac{1}{\pi} \;\Gamma_{N}/2} 
{(\omega\!+\!E_{d})^{2} 
+(\Gamma_{N}/2)^{2}} 
\label{dos_free} 
\end{eqnarray} 
with the quasiparticle energy $E_{d}=\sqrt{\varepsilon_{d}^{2}+
\Delta_{d}^{2}}$. The subgap spectrum consists thus of the particle 
and hole peaks at $\omega\!=\!\pm E_{d}$ whose spectral weights 
depend on $\varepsilon_{d}$ and a broadening is controlled by 
$\Gamma_{N}$. These particle and hole Lorentzians are well separated 
from each other until $\Gamma_{S} \geq \Gamma_{N}$. Otherwise
they merge into a single structure (see the second reference of 
\cite{Domanski-EOM}).

\begin{figure}
\epsfxsize=10cm\centerline{\epsffile{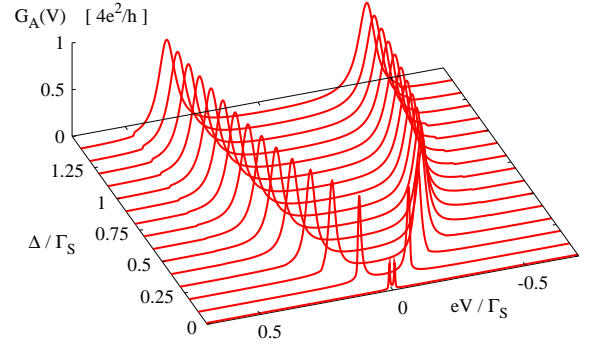}}
\vspace{0.0cm}
\caption{(color online) The subgap Andreev conductance obtained for 
the uncorrelated quantum dot with $\varepsilon_{d}=0$ and asymmetric 
couplings $\Gamma_{N}/\Gamma_{S}=0.1$.}
\label{Andreev_conductance}
\end{figure}

\subsection{Tunneling spectroscopy}

Any experimental verification of the subgap states is possible only
indirectly, by measuring the differential conductance of the tunneling 
current $I(V)$. In general the charge transport induced through the N-QD-S 
junction consists of the quasiparticle (QP) and Andreev (A) currents 
$I(V)=I_{QP}(V)+I_{A}(V)$. They can expressed in the Landauer-type 
form \cite{Krawiec-04}
\begin{eqnarray}
I_{QP}(V) & = & \frac{2e}{h} \int d\omega T_{QP}(\omega) 
\left[ f(\omega-eV) - f(\omega)\right]  
\label{currentQD} \\
I_{A}(V) & = & \frac{2e}{h} \int d\omega T_{A}(\omega) 
\left[ f(\omega-eV)-f(\omega+eV)\right]  
\label{currentA}
\end{eqnarray}
with the Fermi distribution $f(\omega)=\left[ \mbox{\rm exp}
(\omega/k_{B}T) +1 \right]^{-1}$. Transmittance of the Andreev 
channel $T_{A}(\omega)$ depends on the off-diagonal part of 
the Green's function
\begin{eqnarray}
 T_{A}(\omega) = \left| \Gamma_{N} \right|^{2} \; \left|
{\mb G}_{d,12}(\omega) \right|^{2}
\label{T_A}
\end{eqnarray}
whereas the effective quasiparticle transmittance $T_{QP}(\omega)$
contains several contributions
\begin{eqnarray}
 T_{QP}(\omega) &=&  \Gamma_{N} \Gamma_{S} \; \left( \left|
{\mb G}_{d,11}(\omega) \right|^{2} + \left| {\mb G}_{d,12}(\omega) \right|^{2}
\right. \nonumber \\ & - & \left.
\frac{\Delta}{\omega} \mbox{\rm Re} \left\{ 
{\mb G}_{d,11}(\omega) {\mb G}^{*}_{d,12}(\omega) \right\} \right) .
\label{T_QP}
\end{eqnarray}
Usually the off-diagonal Green's function ${\mb G}_{d,12}(\omega)$ quickly
vanishes outside the energy gap, therefore for $|\omega| \geq \Delta$
the tunneling current  simplifies to the popular Meir-Weingreen formula
\begin{eqnarray}
\lim_{|eV|\geq \Delta} I(V) & \approx & \frac{2e}{h} \int d\omega 
\Gamma_{N} \Gamma_{S} \;  \left| {\mb G}_{d,11}(\omega) \right|^{2}
\nonumber \\ & \times &
\left[ f(\omega-eV)-f(\omega) \right] .
\label{current_above_gap}
\end{eqnarray}
In the subgap regime $|\omega| < \Delta_{d}$ (especially for the strongly 
asymmetric couplings $\Gamma_{S} \gg \Gamma_{N}$) the transport is solely
provided by the Andreev current (\ref{currentA}). Figure \ref{Andreev_conductance} 
shows the Andreev conductance $G_{A}(V)=\frac{d}{dV}I_{A}(V)$ obtained for 
$\Gamma_{N}=0.1\Gamma_{S}$. We can notice that the differential conductance 
is similar (although not identical) with the in-gap spectrum $\rho_{d}(\omega)$
presented in figures  \ref{proximity_effect} and \ref{in-gap-energies}. 

\begin{figure}
\epsfxsize=11cm\centerline{\epsffile{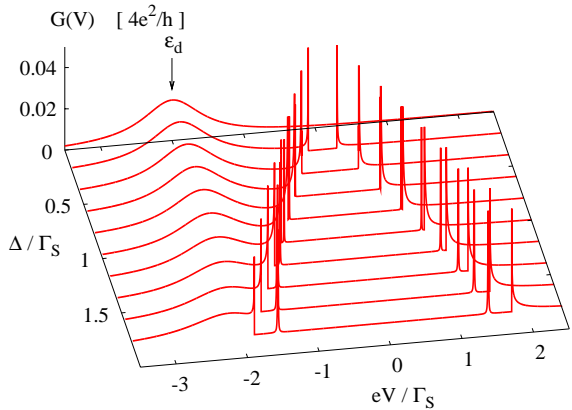}}
\vspace{0.0cm}
\caption{(color online) The effective differential conductance of the 
uncorrelated quantum dot asymmetrically coupled to the external leads 
$\Gamma_{N}/\Gamma_{S}=0.01$ obtained for $\varepsilon_{d}=-2\Gamma_{S}$.}
\label{total_conductance}
\end{figure}

The next plot \ref{total_conductance} illustrates the total conductance 
$G(V)=\frac{d}{dV}I(V)$ obtained at $T=0$ for $\varepsilon_{d}=-2\Gamma_{S}$. 
In these curves we can clearly identify: 
a) the broad peak at $eV = \varepsilon_{d}$,
b) the signatures of gap edge singularities (manifested by sharp enhancement 
   of the single particle tunneling at $eV=\pm\Delta$), and 
c) the well pronounced in-gap features related to the Andreev bound states.
For $\Delta > \Gamma_{S}$ the in-gap features are well separated from 
the gap edge singularities, otherwise it is rather difficult to recognize 
them for the coupling $\Gamma_{N} \geq \Gamma_{S}$  (for instance, see 
figure 2 in the Ref.\ \cite{Koerting-10}). The experimental data of Deacon 
{\em et al} \cite{Deacon-10} clearly confirmed such well defined subgap 
peaks in the Andreev conductance for the strongly asymmetric coupling 
$\Gamma_{S}\geq 40 \Gamma_{N}$. 

\section{Correlation effects}

Interplay between the Coulomb repulsion and the induced on-dot pairing 
is, in general, a very complicated issue. To gain some insight regarding 
their competition  we shall consider the strongly asymmetric limit 
$\Gamma_{S}\gg\Gamma_{N}$ (i.e.\ assuming $\Gamma_{N}=0^{+}$). 
In absence of the Coulomb repulsion the strong hybridization to 
superconducting electrode converts the quantum dot into 'superconducting 
impurity' with the induced pairing gap $\Delta_{d}=\Gamma_{S}/2$. 
Influence of the Coulomb repulsion $U_{d}$ on the subgap Andreev states 
in the large gap limit with a vanishing coupling to the normal lead 
has been first addressed by E.\ Vecino {\em et al} \cite{Vecino-03}.
Comparison of the methods used for determination of the bound states 
of the 'superconducting Anderson model' have been recently revisited 
in Ref.\ \cite{Rodero-12}. 
In what follows below, we briefly summarize the essential results 
based on the exact solution of the effective 'superconducting' QD 
Hamiltonian \cite{Tanaka-07}
\begin{eqnarray}
\hat{H}_{QD} = \sum_{\sigma} \varepsilon_{d} \hat{d}^{\dagger}_{\sigma}
\hat{d}_{\sigma} - \Delta_{d} \left(  \hat{d}^{\dagger}_{\uparrow}
\hat{d}^{\dagger}_{\downarrow} + \hat{d}_{\downarrow}\hat{d}_{\uparrow}\right)
+U \hat{n}_{d\downarrow} \hat{n}_{d\uparrow},
\label{eq23}
\end{eqnarray}
where the proximity effect is taken into account by the pair source/sink terms.
The doublet configurations $\left| \uparrow\right>$ and $\left| \downarrow
\right>$ (corresponding to total spin $S=\frac{1}{2}$) represent  true 
eigenstates with the eigenvalue $\varepsilon_{d}$. The other singlet states 
($S=0$) can be expressed as linear combinations of the empty and doubly 
occupied sites
\begin{eqnarray}
\left| \Psi_{-} \right> & = & u_{d} \left| 0 \right> - v_{d} \left|
\uparrow \downarrow \right> ,\\
\left| \Psi_{+} \right> & = & v_{d} \left| 0 \right> + u_{d} \left|
\uparrow \downarrow \right> .
\end{eqnarray}
The corresponding eigenenergies are given by \cite{Bauer-07,Vecino-03,Wysokinski-12}
\begin{eqnarray}
E_{\mp}=\left( \varepsilon_{d}+\frac{U_{d}}{2} \right) \mp
\sqrt{\left( \varepsilon_{d}+\frac{U_{d}}{2} \right)^{2}+\Delta_{d}^{2}} 
\label{qp_energies}
\end{eqnarray}
and the diagonalization coefficients $u_{d}$, $v_{d}$ take the form
\begin{eqnarray}
u_{d}^{2} = \frac{1}{2} \left[ 1 + \frac{\varepsilon_{d}+U_{d}/2}
{E_{d}} \right] =1 -v_{d}^{2}
\end{eqnarray}
with $E_{d}=\sqrt{\left(\varepsilon_{d}+U_{d}/2\right)^{2}+
\Delta_{d}^{2}}$.
Using the spectral Lehmann representation we can determine the full 
matrix Green's function ${\mb G}_{QD}(\omega)$ of the 'superconducting 
atomic limit' (in the case $\Gamma_{N}\!=\!0^{+}$). Because of the Coulomb
blockade it  takes effectively the four-pole structure
\begin{eqnarray}
{\mb G}_{QD,11}(\omega) &=&   \frac{\alpha\;u_{d}^{2}}
{\omega-\left( \frac{U_{d}}{2}+E_{d} \right)} + \frac{\beta \; 
v_{d}^{2}}{\omega-\left( \frac{U_{d}}{2}-E_{d} \right)} 
\nonumber \\
&+& \frac{\alpha 
\; v_{d}^{2}}{\omega+\left( \frac{U_{d}}{2}+E_{d} \right)} 
+ \frac{\beta \; u_{d}^{2}}{\omega+\left( \frac{U_{d}}{2}-E_{d} \right)}
 \label{G11_atomic} \\
{\mb G}_{QD,12}(\omega) &=&   \frac{\alpha\;u_{d}v_{d}}
{\omega-\left( \frac{U_{d}}{2}+E_{d} \right)}  - \frac{\beta \; 
u_{d}v_{d}}{\omega-\left( \frac{U_{d}}{2}-E_{d} \right)} 
\nonumber \\ & - &
 \frac{\alpha 
\; u_{d}v_{d}}{\omega+\left( \frac{U_{d}}{2}+E_{d} \right)} 
+ \frac{\beta \; u_{d}v_{d}}{\omega+\left( \frac{U_{d}}{2}-E_{d} \right)}
 \label{G12_atomic}
\end{eqnarray}
and ${\mb G}_{QD,22}(\omega) = -\left[ {\mb G}_{QD,11}(-\omega)\right]^{*}$,
${\mb G}_{QD,12}(\omega) = \left[ {\mb G}_{QD,21}(-\omega)\right]^{*}$. 
The relative spectral weights $\alpha$ and $\beta$ are given by
\begin{eqnarray}
\alpha &=& \frac{\mbox{\rm exp}\left\{ {\frac{U_{d}}{2k_{B}T}}\right\}
+\mbox{\rm exp}\left\{ -\; \frac{E_{d}}{k_{B}T}\right\}}
{2\;\mbox{\rm exp}\left\{ {\frac{U_{d}}{2k_{B}T}}\right\}
+\mbox{\rm exp}\left\{ -\; \frac{E_{d}}{k_{B}T}\right\}
+\mbox{\rm exp}\left\{ \frac{E_{d}}{k_{B}T}\right\}}
\nonumber \\
&=& 1 - \beta .
\end{eqnarray}
Spectrum of the correlated quantum dot consists of the four in-gap resonances 
at  quasiparticle energies $\pm \frac{U_{d}}{2} \pm E_{d}$ . For arbitrary 
$U_{d}$ the spectral function $\rho_{QD}(\omega) \equiv -\;\frac{1}{\pi} 
\mbox{\rm Im}\left\{ {\mb G}_{QD,11}(\omega) \right\}$  takes the following form
\begin{eqnarray}
&&\rho_{QD}(\omega) = 
 \label{local_spectrum}  \\
&=&   \alpha u_{d}^{2} \; \delta
\left( \omega- \frac{U_{d}}{2}-E_{d} \right)  
+ \beta v_{d}^{2} \; \delta \left( \omega- \frac{U_{d}}{2}+E_{d} \right) 
\nonumber \\
&+& \alpha v_{d}^{2} \; \delta \left( \omega+ \frac{U_{d}}{2}+E_{d} \right) 
+ \beta u_{d}^{2} \; \delta \left( \omega+ \frac{U_{d}}{2}-E_{d} \right) .
\nonumber
\end{eqnarray}
This spectral function (\ref{local_spectrum}) obeys the sum rule 
$\int_{-\infty}^{\infty} \rho_{QD}(\omega)\; d\omega =1$. For 
$U_{d}\!=\!0$ it properly reproduces the exact BCS-type solution
$\lim_{U_{d}\rightarrow 0} \rho_{QD}(\omega)=u_{d}^{2} 
\delta( \omega - \sqrt{\varepsilon_{d}^{2}+\Delta_{d}^{2}}) 
+ v_{d}^{2} \delta( \omega + \sqrt{\varepsilon_{d}^{2}+
\Delta_{d}^{2}})$. Using  (\ref{eq23}) we can discuss the 
qualitative effects due to a competition between the Coulomb 
interactions and the proximity induced on-dot pairing. This
aspect has been practically investigated in various nanoscopic 
setups \cite{Deacon-10,Aguado-13,Viewpoint}. Expansions around 
this 'superconducting atomic limit' (for $\Gamma_{N}\!\neq\! 0$) 
have been developed in the Refs \cite{Meng-09,Konig_etal}.

\begin{figure}
\epsfxsize=10cm\centerline{\epsffile{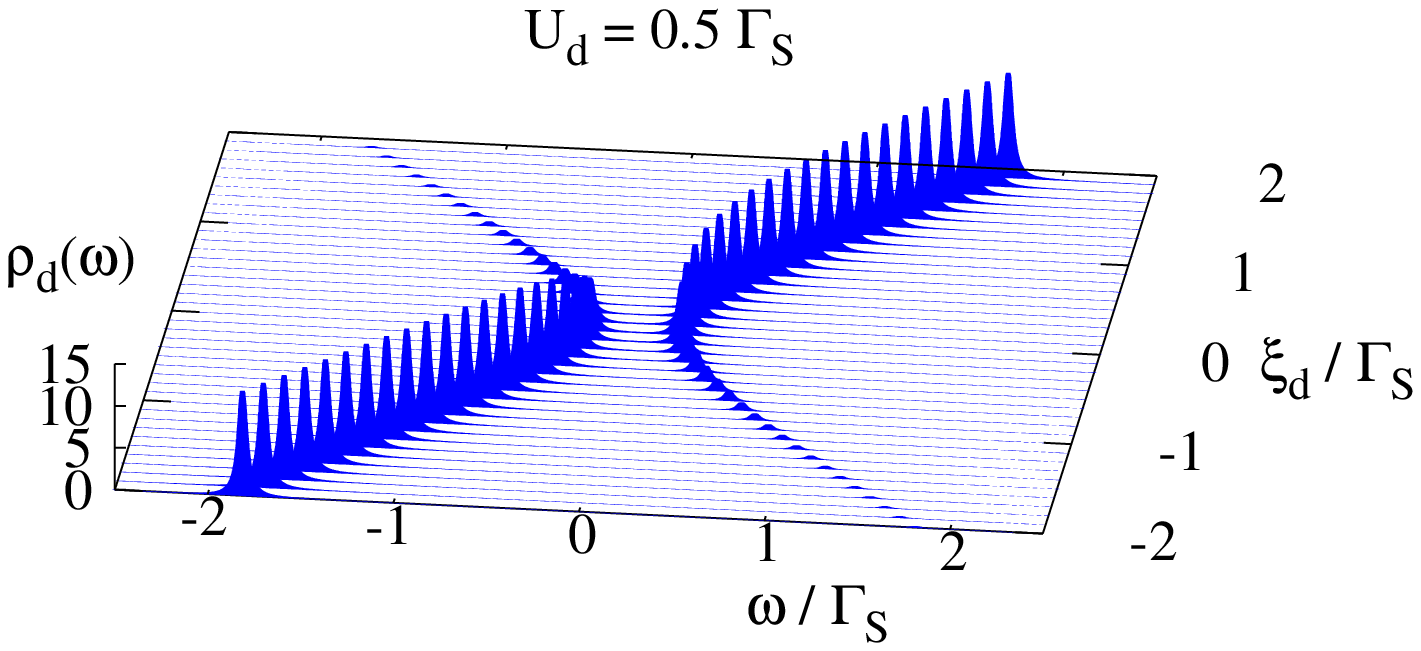}}
\vspace{-1.0cm}
\epsfxsize=10cm\centerline{\epsffile{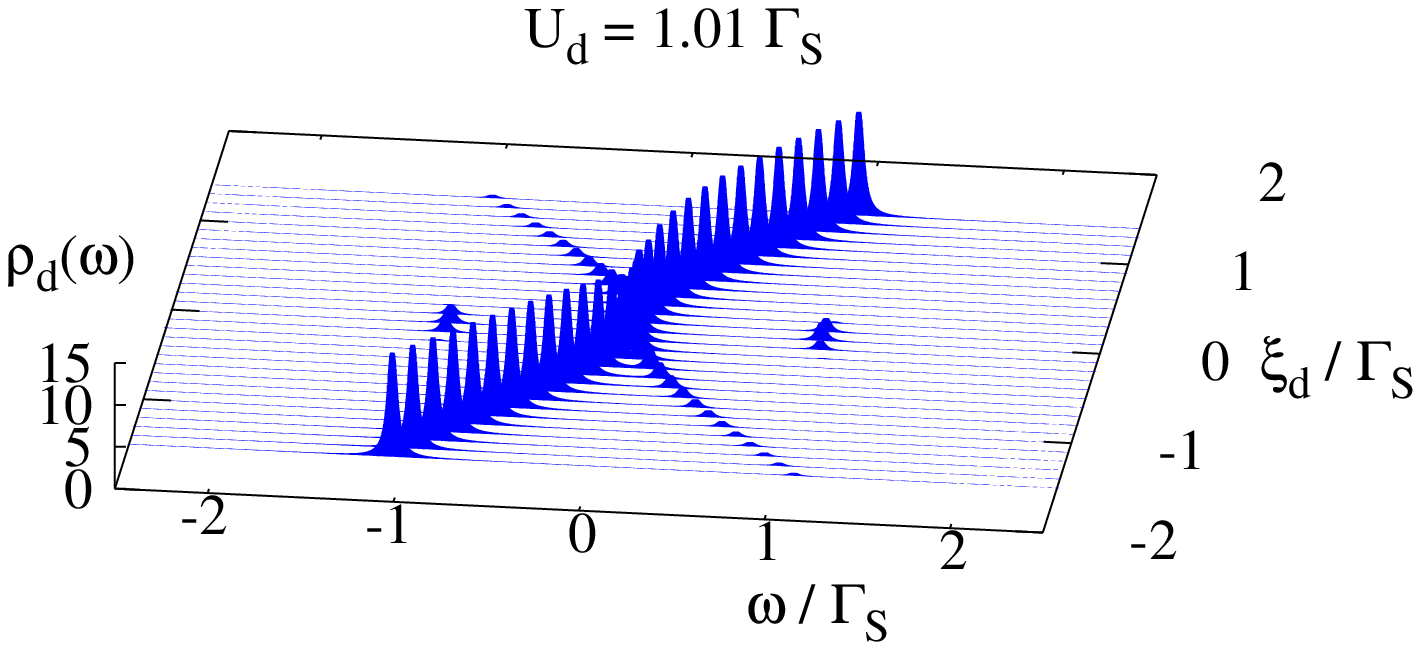}}
\vspace{-1.0cm}
\epsfxsize=10cm\centerline{\epsffile{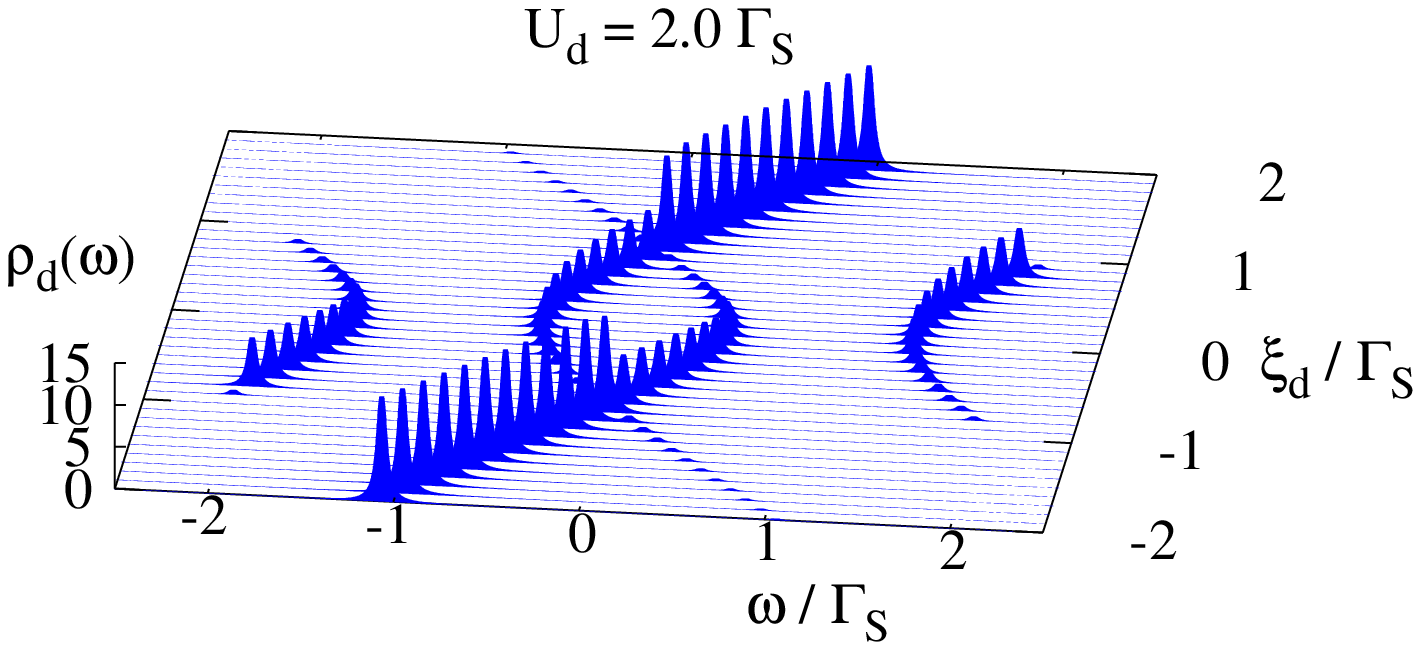}}
\vspace{0.0cm}
\caption{(color online) Spectral function $\rho_{d}(\omega)$ of 
the correlated quantum obtained at $T=0$ in the 'superconducting atomic 
limit' $\Delta \gg \Gamma_S$ 
for several values of the Coulomb potential $U_{d}$. 
We plot the spectral function  with respect to the energy $\xi_{d} \equiv 
\varepsilon_{d}+\frac{U_{d}}{2}$ for $U_{d}/\Gamma_{S}=0.5$ (upper 
panel), $1.01$ (middle panel) and $2.0$ (lower panel).}
\label{spectrum_sc_atomic_limit}
\end{figure}
\begin{figure}
\epsfxsize=10cm\centerline{\epsffile{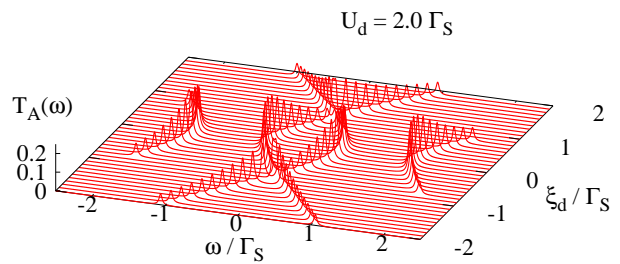}}
\caption{(color online)  Andreev transmittance (\ref{T_A}) obtained 
at $T=0$ in the 'superconducting atomic limit' for $U_{d}=2\Gamma_{S}$.}
\label{TA_atomic}
\end{figure}

Since the most profound influence of the Coulomb repulsion takes place
in the particle-hole symmetric case (i.e.\ for $\varepsilon_{d}=-U_{d}/2$) 
we shall explore this situation, focusing on the fate in-gap resonances  
upon varying $U_{d}$. The Coulomb potential $U_{d}$ directly affects 
the quasiparticle energies $\pm U_{d}/2 \pm \Delta_{d}$ and their 
spectral weights. By inspecting (\ref{qp_energies}) we can notice, 
that for $\Delta_{d}=\frac{1}{2}U_{d}$ the ground state evolves 
from the BCS singlet to the doublet configuration \cite{Bauer-07}. 
This crossover is accompanied by significant redistribution of 
the spectral weights 
\begin{eqnarray}
\lim_{T \rightarrow 0} \alpha = \left\{
\begin{array}{ll}
0 \hspace{0.5cm}& \mbox{\rm for  } \frac{1}{2}U_{d} < \Delta , \\
\frac{1}{3} & \mbox{\rm for  } \frac{1}{2}U_{d} = \Delta , \\
\frac{1}{2} & \mbox{\rm for  } \frac{1}{2}U_{d} > \Delta . 
\end{array}
\right.
\end{eqnarray} 
In the weak interaction limit $\frac{1}{2}U_{d} < \Delta$ (corresponding
to the BCS singlet state) the spectrum consists of two peaks 
\begin{eqnarray}
\rho_{QD}(\omega) &=&   \frac{1}{2} \delta \left( \omega- 
\frac{U_{d}}{2}+\Delta_{d} \right)  + \frac{1}{2} \delta 
\left( \omega+ \frac{U_{d}}{2}-\Delta_{d} \right) ,
\nonumber \\ &&
 \label{BCS_singlet} 
\end{eqnarray}
separated by an effective energy gap $2\Delta_{d}-U_{d}$. At 
the singlet-doublet crossover (i.e.\ for $\frac{1}{2}U_{d} = \Delta$) 
the spectrum evolves to three-pole structure
\begin{eqnarray}
\rho_{QD}(\omega) &=&   \frac{1}{6} \delta \left( \omega 
+2\Delta_{d} \right)  + \frac{1}{6} \delta \left( \omega
- 2\Delta_{d} \right) + \frac{2}{3} \delta \left( \omega
 \right) .
\nonumber \\ &&
 \label{crossover} 
\end{eqnarray}
Finally, in the strong interaction limit $\frac{1}{2}U_{d} > \Delta$
(corresponding to the doublet configuration), the spectral function 
$\rho_{QD}(\omega)$ consists of the four in-gap resonances
\begin{eqnarray}
\rho_{QD}(\omega) &=&   \frac{1}{4} \delta \left( \omega \pm 
\frac{U_{d}}{2}\pm\Delta_{d} \right)  .
 \label{doublet} 
\end{eqnarray}

In the non-symmetric case $\varepsilon_{d}\neq -U_{d}/2$ the 
singlet-doublet crossover occurs at larger values of $U_{d}$, but 
still the QP spectrum comprises either two, three, or four subgap 
Andreev states. We illustrate this behavior in figures 
\ref{spectrum_sc_atomic_limit} and \ref{TA_atomic},
where the coupling $\Gamma_{N}$ to the normal electrode is treated 
via the simple substitution ${\mb G}^{-1}_{d}(\omega) = 
\left[ {\mb G}_{QD}(\omega) \right]^{-1} + \;\frac{i}{2}
\Gamma_{N} \; \mb{I}$ \cite{Wysokinski-12}. Influence of 
the metallic lead  causes then a broadening of the subgap 
states.

\section{Subgap states in the Kondo regime}

In the last section we  address additional effects characteristic 
for the Kondo regime. This issue has been studied in the literature 
by a number of authors using a variety of methods, such as:
equation of motion technique \cite{Fazio-98},
slave boson approach \cite{{Raimondi-99},Krawiec-04}, 
non-crossing approximation \cite{Clerk-00}, 
iterated perturbation theory \cite{Cuevas-01,Yamada-11}, 
Keldysh Green's function approach combined with the path
integral formalism and dynamical mean field approximation \cite{Avishai-01}, 
numerical renormalization group \cite{Tanaka-07,Bauer-13}, 
modified equation of motion approach \cite{Domanski-EOM},
functional renormalization group \cite{Karrasch-08},
cotunneling approach (for the spinful dot) \cite{Koerting-10},
Quantum Monte Carlo simulations \cite{Koga-13} and other 
\cite{Kang-98,Sun-99}. In particular, it has been checked whether 
the Kondo resonance could be somehow manifested in the subgap conductance.

For studying the qualitative features caused by the Kondo effect 
we shall determine the total selfenergy ${\mb \Sigma}_{d}(\omega)
={\mb \Sigma}^{0}_{d}(\omega)+{\mb \Sigma}^{U}_{d}(\omega)$
in the matrix form
\begin{eqnarray} 
{\mb \Sigma}_{d}(\omega) = 
\left( \begin{array}{cc}
\Sigma^{diag}_{\uparrow}(\omega) & \Sigma^{off}(\omega) \\ 
\left[ \Sigma^{off}(-\omega) \right]^{*} & 
-\left[ \Sigma^{diag}_{\downarrow}(-\omega) \right]^{*}
\end{array} \right) 
\label{U_diagonal} 
\end{eqnarray} 
focusing on the subgap limit $\Delta \gg \Gamma_{S}$. We thus 
consider the correlated quantum dot with the induced on-dot 
pairing $\Delta_{d}$ coupled to the metallic lead
$\hat{H} = \sum_{\sigma} \varepsilon_{d} \hat{d}^{\dagger}_{\sigma}
\hat{d}_{\sigma} +U \hat{n}_{d\downarrow} \hat{n}_{d\uparrow} 
- \Delta_{d} \left(  \hat{d}^{\dagger}_{\uparrow}
\hat{d}^{\dagger}_{\downarrow} + \hat{d}_{\downarrow}
\hat{d}_{\uparrow}\right)+\hat{H}_{N}+\sum_{{\bf k},\sigma } 
\left( V_{{\bf k} N} \; \hat{d}_{\sigma}^{\dagger} \hat{c}_{{\bf k} 
\sigma N } + \mbox{\rm h.c.} \right)$. This simplified problem is 
not solvable exactly therefore we have to impose some approximations.
For this purpose we treat the Coulomb interactions within the selfconsistent
 scheme based on the equation of motion (EOM) approach \cite{EOM} 
extended to the case of the on-dot pairing $\Delta_{d}\neq 0$. 

As a starting point we express a diagonal part of the Green's 
function ${\mb G}$ by the  BCS-type pairing Ansatz
\cite{Levin_etal}
\begin{eqnarray} 
\left[ {\mb G}_{11}(\omega) \right]^{-1} &=& 
\omega-\varepsilon_{d}-\Sigma^{diag}_{\uparrow}(\omega) \nonumber \\
&-&\frac{\Delta_{d}^{2}}{\omega+\varepsilon_{d}+
\left[\Sigma^{diag}_{\downarrow}(-\omega)\right]^{*}} .
\label{BCS_ansatz}
\end{eqnarray} 
The hole propagator is related to (\ref{BCS_ansatz}) via 
${\mb G}_{22}(\omega) = -\left[ {\mb G}_{11}(-\omega)\right]^{*}$.
Let us notice that in the noninteracting case 
$\lim_{U_{d}\rightarrow 0} \Sigma^{diag}_{\sigma}(\omega) =
-i\Gamma_{N}/2$. For arbitrary $U_{d}\neq 0$ we estimate 
$\Sigma^{diag}_{\sigma}(\omega)$ using  the  EOM decoupling 
procedure \cite{EOM} (for details see Appendix B in Ref.\ \cite{Baranski-11})
but our scheme outlined below can be combined also with other approximations, 
for instance NCA \cite{Clerk-00}, perturbative expansion \cite{Cuevas-01} 
etc. The EOM approach yields  
\begin{widetext}
\begin{eqnarray} 
\Sigma^{diag}_{\sigma}(\omega)  \simeq U_{d} \left[ 
n_{d,\bar{\sigma}}\!-\!\Sigma_{1}(\omega)\right] 
+ \frac{U_{d} \left[ n_{d,\bar{\sigma}}\!-\!
\Sigma_{1}(\omega)\right] \left[ \Sigma_{3}(\omega)
+U_{d}(1\!-\!n_{d,\bar{\sigma}})\right]}{\omega
-\varepsilon_{d}-\Sigma_{0}(\omega)-\left[ 
\Sigma_{3}(\omega)+U_{d}(1-n_{d,\bar{\sigma}})\right]} ,
\label{sigma_EOM}
\end{eqnarray} 
where $\Sigma_{0}(\omega)=-\frac{i}{2}\Gamma_{N}$,
\begin{eqnarray}
\Sigma_{\nu}(\omega) = \sum_{{\bf k}} 
\! |V_{{\bf k} N}|^{2} \!\! \left[ \frac{1}{\omega\! - \! 
\xi_{{\bf k} N} } + \frac{1}{\omega\! -\! U_{d} \! - 
2 \varepsilon_{d}\!+\!\xi_{{\bf k} N} } \right]  
\; \times \; \left\{ \begin{array}{lc} 
f(\xi_{{\bf k}N}) & \mbox{\rm for } \nu=1 \\
1 & \mbox{\rm for } \nu=3   \end{array}
\right.  
\label{sigmas} 
\end{eqnarray}
\end{widetext}
and $\bar{\uparrow}=\downarrow$, $\bar{\downarrow}=\uparrow$.
To determine the off-diagonal parts of ${\mb G}(\omega)$ we next
use the  following exact relation
\begin{eqnarray} 
\left( \omega - \varepsilon_{d} \right) {\mb G}_{11}(\omega) 
= 1 + U_{d} \langle\langle \hat{d}_{\uparrow}\hat{n}_{d\downarrow};
\hat{d}_{\uparrow}^{\dagger}\rangle\rangle - \Delta_{d} {\mb G}_{12}(\omega)
\label{exact_relation}
\end{eqnarray} 
and  ${\mb G}_{21}(\omega)={\mb G}_{12}^{*}(-\omega)$. As an approximation 
we neglect here an influence of the induced on-dot pairing on the two-body 
propagator $\langle\langle \hat{d}_{\uparrow}\hat{n}_{d\downarrow};
\hat{d}_{\uparrow}^{\dagger}\rangle\rangle$ appearing in (\ref{exact_relation}).
This assumption should be justified as long as $U_{d}$ is safely 
larger than $\Delta_{d}=\Gamma_{S}/2$. We thus  take 
\begin{eqnarray} 
\langle\langle \hat{d}_{\uparrow}\hat{n}_{d\downarrow};
\hat{d}_{\uparrow}^{\dagger}\rangle\rangle \simeq \frac{ 
n_{d\downarrow}  - \Sigma_{1}(\omega) \; 
{\mb G}_{11}(\omega)}{ \omega - \varepsilon_{d} 
-\Sigma_{0}(\omega) - U_{d} - \Sigma_{3}(\omega)} 
\label{expression_for_G2}
\end{eqnarray} 
which formally originates from the EOM solution \cite{EOM}.

Having this first guess for the matrix Green's function ${\mb G}(\omega)$ 
[expressed through the equations (\ref{BCS_ansatz}-\ref{expression_for_G2})] 
we now construct its selfconsistent improvement. We update the initial pairing 
Ansatz (\ref{BCS_ansatz}) by iteratively substituting the former 
selfenergy functional ${\mb \Sigma}[{\mb G}(\omega)]$ 
to the true relation
\begin{eqnarray} 
\left[ {\mb G}_{11}(\omega) \right]^{-1} &=& 
\omega-\varepsilon_{d}-\Sigma^{diag}(\omega) \nonumber \\
&-&\frac{\left[{\mb \Sigma}^{off}(-\omega)
\right]^{*}{\mb \Sigma}^{off}(\omega)}
{\omega+\varepsilon_{d}+
\left[\Sigma^{diag}(-\omega)\right]^{*}} . 
\label{true_relation}
\end{eqnarray} 
At each step we determine the off-diagonal terms via (\ref{exact_relation}) 
and continue until a satisfactory convergence is reached. We have done 
numerical calculations of the matrix Green's function ${\mb G}(\omega)$  
within such algorithm using the mash of 9000 equidistant energies 
$\omega_{n}$  slightly above the real axis. We have noticed that 
practically 7 to 11 iterations were sufficient for a good convergence.

Let us remark that upon neglecting the terms 
$\Sigma_{1}(\omega)$ and $\Sigma_{3}(\omega)$ of the diagonal 
selfenergy (\ref{sigma_EOM}) in the initial iterative step we would 
recover the usual second order perturbation formula 
$\lim_{\Sigma_{1},\Sigma_{3} \rightarrow 0} 
\Sigma^{diag}_{\sigma}(\omega)  = U_{d}
n_{d,\bar{\sigma}} \label{self_comp} + U_{d}^{2} \; 
\frac{ n_{d,\bar{\sigma}} (1\!-\!n_{d,\bar{\sigma}})}
{\omega+i\Gamma_{N}/2-\varepsilon_{d}-U_{d}(1-n_{d,\bar{\sigma}})}$.
This fact indicates that such simplified selfenergy is able to account 
for the charging effect (i.e.\ the Coulomb blockade) discussed 
by us in the preceding section. In our numerical treatment we keep, 
however, all the contributions entering  (\ref{sigma_EOM}) because they 
are important in the Kondo regime $\varepsilon_{d}<0<\varepsilon_{d}+U_{d}$. 
At temperatures below $k_{B}T_{K}=0.5
\sqrt{U_{d}\Gamma_{N}} \mbox{\rm{exp}} \left\{ -\pi \frac{| 
\varepsilon_{d} \left( \varepsilon_{d}+U_{d}\right) |}{U_{d}\Gamma_{N}} 
\right\}$ the diverging real part of $\Sigma_{1}(\omega)$ induces 
then a narrow Abrikosov-Suhl (or Kondo) peak at $\mu_{N}$. From more 
sophisticated treatments it is known that at low temperatures its 
broadening should scale with $k_{B}T_{K}$. Unfortunately the EOM 
approach does not reproduce the low energy structure of the Kondo 
peak. This missing information could be obtained e.g.\ from the 
renormalization group calculations \cite{Tanaka-07,Bauer-13}, but
we nevertheless hope that the overall spectrum and the transport 
properties are qualitatively properly reproduced by the present 
treatment.  

\begin{figure}
\epsfxsize=10cm\centerline{\epsffile{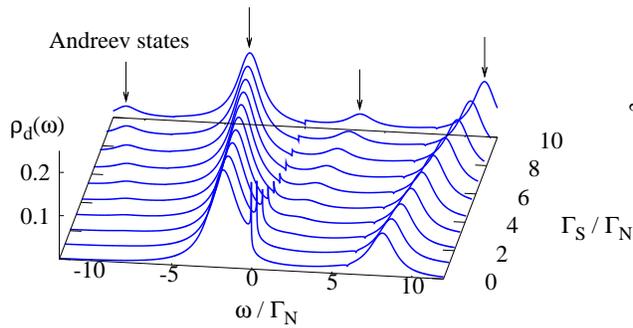}}
\vspace{0.0cm}
\caption{(color online) Spectral function $\rho_{d}(\omega)$ of the strongly 
correlated quantum dot obtained in the Kondo regime for $\varepsilon_{d}=
-2\Gamma_{N}$, $U_{d}=10\Gamma_{N}$ at temperature $T=0.001\Gamma_{N}
k_{B}^{-1}$ (well below $T_{K}$). Calculations have been done assuming 
$\Delta$ much larger than $U_{d}$.}
\label{spectrum_in_Kondo_regime}
\end{figure}

In figure \ref{spectrum_in_Kondo_regime} we show the spectral function
of the strongly correlated quantum dot obtained for $\varepsilon_{d}=
-2\Gamma_{N}$, $U_{d}=10\Gamma_{N}$ at $k_{B}T=0.0001\Gamma_{N}$.
The curve corresponding to $\Gamma_{S}=0$ (in absence of the proximity 
effect) reveals the quasiparticle peak at $\varepsilon_{d}$ 
and its Coulomb satellite at $\varepsilon_{d}+U_{d}$. Both peaks 
are broadened by $\sim \Gamma_{N}$ (actually, the EOM approximation 
 slightly overestimates such broadening). Besides the quasiparticle 
in-gap states we also notice the narrow Kondo resonance at $\omega=0$.

\begin{figure}
\epsfxsize=10cm\centerline{\epsffile{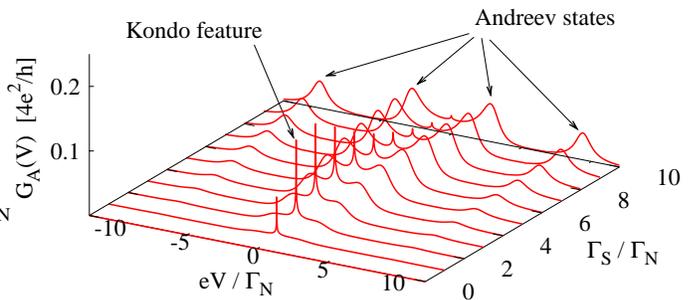}}
\vspace{0.0cm}
\caption{(color online) The differential Andreev conductance $G_{A}(V)$ 
of the Kondo regime obtained for the same model parameters as in figure
\ref{spectrum_in_Kondo_regime}. The zero-bias enhancement (caused by 
the Abrikosov-Suhl resonance) is gradually suppressed for increasing
$\Gamma_{S}$ due to the on-dot pairing.}
\label{GA_in_Kondo_regime}
\end{figure}

For increasing values of the coupling $\Gamma_{S}$ there occurs
a gradual formation of the particle and hole in-gap features 
(similar to what has been discussed in the previous sections).
This process is accompanied by some qualitative changes of 
the Kondo resonance. It is gradually suppressed and, for 
$\Gamma_{S}\geq4\Gamma_{N}$ evolves to the kink-type 
structure characteristic for the mixed valence regime. 
This behavior is caused by a competition between the on-dot pairing 
(promoted by $\Gamma_{S}$) and the Kondo-type correlations \cite{Glazman-89}. 
Signatures of their eventual coexistence occur when $\Gamma_{S}$ is comparable 
or slightly larger than $\Gamma_{N}$. Under such conditions the subgap 
spectrum consists of four Andreev quasiparticle peaks (of 
a broadening $\sim \Gamma_{N}$) and the
narrow Kondo resonance. These in-gap features can be practically
observed by the measurements of the differential conductance in 
the Kondo regime. The subgap conductance shown in figure 
\ref{GA_in_Kondo_regime} reveals all the mentioned peaks  
of the electronic spectrum (see figure \ref{spectrum_in_Kondo_regime})
although in a symmetrized way $G_{A}(V)=G_{A}(-V)$ because the particle
and hole degrees of freedom equally participate in the Andreev
scattering.

\section{Conclusions}

We have investigated the spectroscopic and transport properties of the quantum 
impurity coupled to one conducting and another superconducting electrode.  
The proximity effect depletes electronic states in a subgap region 
$-\Delta < \omega < \Delta$ but the Andreev-type scattering 
(i.e.\ conversion of electrons into the Cooper pairs with a simultaneous 
reflection of the holes) contributes the in-gap states, both in the 
correlated and the uncorrelated quantum dots. Since the Andreev 
mechanism involves the particle and hole degrees of freedom there
appears an even number of the in-gap bound states. 

For a weak coupling to the metallic lead $\Gamma_{N} \ll \Gamma_{S}$ 
the in-gap states take a form of the narrow resonances, representing 
the infinite life-time quasiparticles. Otherwise, the in-gap states 
acquire a finite line-broadening, roughly proportional to $\Gamma_{N}$.
The number of in-gap states depends sensitively on the Coulomb potential
$U_{d}$ (strictly speaking on the ratio $U_{d}/\Delta$). For the weak 
interaction limit the ground state is preferred as the BCS singlet 
configuration \cite{Bauer-07,Meng-09,Konig_etal}, so consequently 
there appear only two Andreev subgap states. For stronger correlations 
$U_{d} \gg \Delta$ the number of in-gap states is doubled. Positions 
of these in-gap states depend on the ratio $\Delta/\Gamma_{S}$. For 
$\Gamma_{S} \leq \Delta$ the in-gap states are located nearby the gap 
edge singularities $\sim \pm \Delta$. Otherwise, they move aside 
from the gap edges and, in the extreme 'superconducting atomic limit'  
$\Gamma_{S}\gg\Delta$, the subgap quasiparticle energies approach 
$\pm \frac{1}{2}U_{d}\pm\sqrt{\left( \varepsilon_{d} + U_{d}/2 
\right)^{2}+\Gamma_{S}^{2}/4}$. Coupling to the metallic lead 
merely affects the line-broadening of such in-gap states.

We have also studied the Kondo regime using iterative scheme based on the equation 
of motion approximation \cite{EOM}. In addition to the previously indicated 
Andreev states we have obtained the narrow Abrikosov-Suhl peak at 
$\omega=0$. Such Kondo feature is present in the spectrum
unless a strong enough coupling to the superconducting lead $\Gamma_{S}$ 
(promoting the on-dot pairing) gradually suppresses it.  
In a region where the Kondo state coexists with the induced on-dot pairing 
the spectral function $\rho_{d}(\omega)$ is characterized by five 
subgap states:  four of them represent the Andreev peaks (with a 
line-broadening $\sim \Gamma_{N}$) and another one is due to the 
Abrikosov-Suhl resonance (of a broadening  $\sim k_{B}T_{K}$). 
These features  show up indirectly in the subgap conductance. 
Signatures of the zero-bias enhancement of the Andreev conductance 
\cite{Deacon-10,DeFranceschi-12,S-QD_and_N-S} can be naturally 
assigned to the Kondo effect. We hope that the present study 
discussing influence of: (i) the asymmetric couplings 
$\Gamma_{N}/\Gamma_{S}$, (ii) the energy gap $\Delta/\Gamma_{S}$ 
and (iii) interplay between the Coulomb repulsion $U_{d}$ and 
the induced on-dot pairing (promoted by $\Gamma_{S}$) 
would be useful for the experimental studies 
of the many-body effects  in the N-QD-S junctions and in their 
more complex multi-terminal equivalents
\cite{Konig_etal,Andergassen-12,LevyYeyati-12,Bauer-13,Viewpoint}.  

\acknowledgments
{We acknowledge discussions with J.\ Bauer, S.\ Andergassen, 
K.I.\ Wysoki\'nski and kindly thank for useful comments from 
R.\ Aguado and Y.\ Avishai.}

\end{document}